# Coherent Activation of Zero-Field Fiske Modes in Arrays of Josephson Junctions

I. Ottaviani, M. Cirillo, M. Lucci, V. Merlo, and M. Salvato[*]

Dipartimento di Fisica and *MINAS Laboratory*
Università di Roma "Tor Vergata"
I-00133 Roma, Italy
[(*)] also at Laboratorio Regionale *Supermat* CNR-INFM
I-84081 Baronissi, Italy

M. G. Castellano and G. Torrioli
IFN- CNR, via Cineto Romano 42
I-00156 Roma, Italy

F. Mueller and T. Weimann

PTB Braunschweig
Bundesallee 100
D-3300 Braunschweig, Germany

## Abstract

Series arrays of Josephson junctions show evidence of a mode in which all the junctions oscillate in synchronism on voltage resonances appearing , in zero external magnetic field, at multiples of the fundamental Fiske step spacing. The measurements show that the current amplitude of the resonances increases linearly as their voltages are summed. Investigation of the nature of the coherent mode by magnetic field responses of arrays and isolated juctions reveals that the oscillations take place in a parameter plane region where dc magnetic fields only activate boundary current and flux-quanta dynamics can take place.

Arrays of Josephson junctions have been subject of noticeable interest over the past decades due the interesting aspect of their dynamics leading to developments both at fundamental and applied physics level [1,2,3]. Coherence of the arrays in the microwave and millimeter-wave range which could provide integrated oscillators for superconducting electronics was investigated in several publications [2,4] but interesting work has more recently been devoted to the study of complex nonlinear excitations [1]. Arrays of long junctions, namely junctions having internal spatial degrees of freedom, operated either in zero applied external field [4,5] or under the influence of an external magnetic field [6], have also been investigated.

The Fiske modes-based dynamics of long Josephson junctions was systematically characterized [7] and a relevant feature of this dynamics is the fact that a threshold dc magnetic field is required for the activation of stable Fiske steps. Below this threshold field complex phenomena and fluxon oscillation-based dynamics may occur [8]. An earlier publication [9] demonstrated, however, that stable Fiske modes can also appear in Josephson junctions in zero-applied magnetic field when the Josephson tunneling current of the junctions is spatially non uniform, a condition that could be generated in the specific case of photo-sensitive junctions. Moreover, Kautz [10] suggested that Fiske modes could be stimulated by a microwave field having a frequency matching that of the Fiske modes $\nu_{FS} = \left(\dfrac{\bar{c}}{2L}\right)$ where $\bar{c}$ is the speed of light in the oxide barrier (typically few percents of the the speed of light in vacuum). Grønbech-Jensen and Cirillo [11] confirmed this sensitivity of Fiske modes to rf fields showing that long junctions operated on the Fiske modes can phase-lock to a boundary external time-varying field for substantial current intervals. In this paper we will show that a peculiar geometry of a Josephson junction and the exposure to self-generated rf fields can trigger an interesting collective mode in a series array of junctions.

Our samples are series arrays arranged in a meander line in a design very similar to that reported in ref. 6 ; the technological process, relying on the basic Nb/Al-AlOx/Nb junction trilayer technology and definition of junction areas by e-beam lithography is described elsewhere [12]. In Fig. 1 we show current-voltage characteristics of three arrays of long inline junctions [13]. For short and long dimensions we herein refer, as usual, to the physical dimensions of the junctions compared to the Josephson penetration depth $\lambda_J = \sqrt{\dfrac{\hbar}{2e\mu_0 d J_c}}$ where $h/2e=\Phi_0=2.07\mathrm{x}10^{-15}$ Wb is the quantum of magnetic flux, $\mu_0 = 4\pi \mathrm{x} 10^{-7}$ H/m, $d$ is the sum of the London penetration depths and thickness of the oxide layer, and $J_c$ is the maximum Josephson pair current density. The long junctions had a symmetric inline geometry [13,14] with only one physical dimension larger than $\lambda_j$. The current-voltage characteristic of the samples shown in Fig. 1 a,b,c are relative to arrays with a different number of junctions and produced in different fabrication batches; in particular for (a) we had a current density of *400*A/cm$^2$ for (b)*190* A/cm$^2$ , and for (c) *470* A/cm$^2$ . The long junctions in (a) had a physical length of *74.5* μm, while those for (b) and (c) had a length of *54.5* μm; the normalized lengths of the junctions for (a), (b) and (c) were respectively $3\lambda_j$, $1.5\lambda_j$ , and $2.3\lambda_j$ ; all the junctions had a width of *4* μm.

Independently upon the number of junctions and current density we can see that the "switching" current distribution appears to be a linear function of the number of junctions, over most of the voltage interval, a peculiarity which is not observed on small Josephson junctions arrays placed on the same chips which show evidence instead of a rather narrow distribution of the Josephson current for all the junctions, as we see in the inset of Fig. 1c. Moreover, we observe that the sample shown in Fig. 1a does not have a superconducting groundplane (which we deposit on the back of the wafers) while those of Fig. 1b,c did have it.

The measurements shown in Fig. 1 were performed at *4.2*K in a liquid helium bath with the samples shielded by a "cold" cryoperm shield , placed at *4.2*K around the samples, and a room temperature mumetal shield surrounding the liquid helium dewar. A quantitative estimate based on the current-voltage (IV) characteristics of Fig. 1 shows that the highest observed value of the "switching currents" corresponds roughly to the maximum value of the Josephson currents of inline junctions calculated on the basis of the Owen and Scalapino model [14]. An evident feature of the IV curves shown in Fig. 1 is that, in spite of the linear slope of the "switching currents", the quasi-particle branches are flat and uniform, meaning that the presence of short circuits on edges or other mechanical defects in the tunneling barriers are not affecting the samples. In conclusion we can say that the phenomenon reported in Fig. 1 does not depend on a specific current density neither on a peculiar geometrical length or on a specific stripline configuration (as we said above the sample of Fig. 1a did not have a groundplane). At this point we can only say that a length of the junctions larger (even slightly) than $\lambda_j$ is necessary in order to observe the effect.

A zoom of the current voltage characteristics of the arrays reveals that the "switching currents" shown in Fig. 1 do not correspond to switches from the maximun Josephson currents of the individual junctions. The reality of the switching dynamics is shown in Fig. 2: this figure shows a sequence of voltage singularities obtained by superimposing a dc voltage offset to the low frequency current sweep that was feeding the junctions for the recording of the IV curves. The switchings from the maximum current of the voltage singularities produce a stairway which, displayed in a compressed scale, generates the linear slope shown in Fig. 1. Since all the resonances have amplitudes larger than the Josephson currents we only see the switching of one zero-voltage current of the array. The voltage spacing of the singularities corresponds to that of the Fiske modes of individual junctions [13]. Indeed, from the position of Fiske modes and zero-field steps [13] in individual "long" and "short" junctions we estimated that the speed of light in the oxide barrier for our samples is $\bar{c} = 0.035c$ and therefore the spacing of the Fiske steps, for a junction long 54.5 μm should be *200* μV like we see in Fig. 2. The depressed value of all the Josephson currents of the array has a strong analogy with the fact that, under a dc magnetic field, the Josephson current is minimum when the first Fiske step attains its maximum value [7,13].

The singularities voltages add up to the sum-gap voltage of the series array meaning that the mode of oscillation at 97 GHz is shared by all the junctions in the array; also, we must bear in mind that we have only two probes for each end of the array and therefore the bias current is feeding the whole array independently upon the number of visualized junctions. The fact that in the IV curves the current amplitude of the singularities increases linearly as the number of junctions must be interpreted in the sense that each displayed singularity (or groups of singularities) with a given amplitude corresponds to have a given number of junctions oscillating in synchronism. The IV curves of Fig. 1 and Fig. 2 tell us that the series array supports a mode for which there is a linear correspondence between the voltage position of the resonances (depending essentially on the number of junctions whose voltage is being added) and their current amplitude. A linear increase of the amplitude of a current as a function of the oscillators involved in a coherent mode is usually interpreted as a signature of a superradiant behavior [15].

We measured the magnetic field dependence of the IV curves of Fig. 1 and the result, for the sample of Fig. 1c is shown in Fig. 3. We see here in (a) that all the currents of the array are uniformly lowered by the external magnetic field and in (b) we report the result of the dependence of the measured (asymptotic) slopes of (a) as a function of the applied magnetic field. We can clearly see in Fig. 3b that the dependence has two distinct tendencies and the first branch has a more pronounced slope. A first physical insight of this change in slope can be gained looking at the data in the inset which show the beginning of the modulations of the first Fiske mode of one junction of the array of Fig. 2a that we biased independently (no current through all the other junctions). We see in the inset that the modulations begin just at *4* gauss, where the slope of the curve changes, meaning that the system is stepping from one dynamical state to another. We recall that the stable Fiske modulations take place when the extenal magnetic field penetrates in the junctions above the value $2\lambda_j J_c$ while below this value complex flux quanta dynamics can take place [7,8]. Note that an estimate of this magnetic field value for the sample of Fig. 2a from the values of $J_c$ and $\lambda_j$ gives *3.0* gauss.

As we just said we were successful in contacting (by a careful on-chip bonding) one single junction of the array of Fig. 2a which could then be dc-biased without feeding all the others. A particular of the current-voltage characteristics of this isolated junction is the one shown in the inset of Fig. 2a where we see the Josephson current and two zero-field steps [13] spaced *400*μV which were maximized by a slight external magnetic field. The stable modulations of the Fiske modes shown in the inset of Fig. 3b were taken from this same junction. Biasing only this junction we do not see evidence in its IV curve of the resonances shown in Fig. 2a which were displayed, in the same measurement run, by feeding current through the whole array.

From the dependence of the slope upon the magnetic field we extract the information that the coherent oscillation is strongly related to flux quanta dynamics. The system, starting from a small start-up "magnetic" signal is generating phase-locked dynamics of the flux-quanta as described, for example, in ref. 18. In that case it was shown that capacitive coupling between the junctions could give rise to the coherent oscillations from an initially disordered spatial distribution of magnetic flux-quanta. Following Fig. 4 we can provide an argument which is strongly in favor of this conjecture : here we show in (a) the IV curve of a small junctions (physical dimensions of the order of *0.2*$\lambda_j$ ) meander array taken on a shielded sample to which we applied by a coil, located inside the cold cryoperm shield, a uniform magnetic field generating "standard" Fiske steps in all the junctions. The Fiske steps have the expected voltage spacing, namely *1.1*mV : in the figure we see only the first Fiske step of each junction along with other IV branches generated by the fact that the Josephson currents are not suppressed. In Fig. 4b instead we see the IV curve of the same array of (a) taken without magnetic shielding : we observe now, even for this small junction array, the same linear increase of the current height of the resonances. The presence of the irregular distribution of magnetic flux along the array generated by the earth magnetic field (roughly *0.5* gauss) can trigger the system on a mode displaying the same features observed for the long junctions. For the long junctions the difference is just that a smaller field is required to start-up the oscillations based on flux-quanta dynamics. Numerical simulations will surely help to understand how the internal dynamics of the junctions is related to the collective mode, however, we believe that the overall picture we have traced so far is quite consistent.

As far as the specific reasons that could give rise to the start-up signal in the long junctions we recall that a work by Camerlingo et al. [9] showed that a Josephson current profile peaked at the edges of the junctions favors the generation of Fiske oscillations, even when no magnetic fields are applied to the junctions. Although the evidence was shown for photo-sensitive junctions [13], the model showed that a mechanism generating pair current peaked at the edges can favor the excitation of the Fiske resonances; in our case, the same inline shape of the Josephson junctions favors the formation of boundary currents. It is also reasonable to suppose that, due to the very limited uncertainty on the definition of the geometrical properties of the junctions (the areas of the junctions are defined by electron beam lithography [12]) exchange of excitations and collective modes relying on geometrical factors are much enhanced. We also remark that, when biased on the current-voltage singularities of Fig.2 the arrays emitted very little radiation (detected by the scheme described in ref. 6). However, above the value of *4* gauss (with the junctions working in the series mode and biased on "stable" Fiske modes) a much larger amount of radiation was coupled into a detector array. The last observation tells us that in the mode shown in Fig. 1 and Fig.2 the arrays employ most of the available energy to sustain internal excitations and leaving tiny amount of power to be coupled out.

In zero applied external field the resonances of Fig. 2a were very stable and could provide voltage sums well above one volt. These resonances, being not vertical like Shapiro steps, would hardly provide voltages useful for metrology purposes [16,17]. We speculate, however, that an injection locking-signal, whose frequency corresponds to a subharmonic of our 97 GHz mode (say in the X-band, for example) by room temperature or "cold" oscillators, could force the resonances to display vertical portions. The resonances, as we see in Fig. 1 and Fig. 2 have current amplitudes going up to several hundred microamperes and, being generated by a collective mode, could survive the dynamical instabilities generated by pumping with low frequency signals [3]. We recall that at the early stages of the voltage standard research the possibility of using "resonant" junctions for the Josephson voltage standard was investigated, but the interaction of the pump signal with Josephson plasma oscillations and internal cavity modes allowed to obtain very limited intervals of stability for the current of the steps. In the present case, the fact that the geometrical properties are defined by electron beam lithography could enhance the collective mode and improve the dynamical stability of the resonances.

In conclusion, we have reported on the evidence of a collective mode in series array of Josephson junctions; the mode is based on oscillations of the individual junctions corresponding to the fundamental Fiske step resonance. The IV curves of the arrays demonstrate that the system is characterized (for most of the displaced voltage interval) by a linear relation between current height of resonances and junction number, typical of superradiant systems. We were able to current-feed one junction independently upon the rest of the array finding that the observed "anomalous" resonances in the IV curves could indeed be obtained only when the whole array was biased. We provided evidence that the start up of the collective mode can be generated by an initial irregular magnetic flux distribution inside the junctions. Finally we have discussed the possible impact of the observed phenomena in fundamental quantum voltage standard research.

**FIGURE CAPTIONS**

Fig. 1  Current-voltage characteristics of long Josephson junctions arrays displaying an anomalous increase of the "switching currents"; in (a), (b) and (c) the arrays contain respectively 352, 374, and 1394 junctions. The inset of (c) shows the very regular characteristics of a small junctions array placed on the same chip.

Fig. 2  Sequences of Fiske steps obtained for a series array respectively of 374 (a) and 1394 (b) junctions. The inset in (a) shows the result of biasing only one junction of the series array without feeding all the others : this junction was long 54.5 μm (2.2 $\lambda_j$ ) and, consistently, we observe two zero field steps spaced 400 μV (twice the spacing of the Fiske modes).

Fig. 3  (a) Dependence of the slopes in the I-V curves shown in Fig. 1c upon the external magnetic field; (b) the slope of the straight lines the Fiske modes in (a) plotted as a function of the external magnetic field. The inset here shows the modulations of stable Fiske steps of the junction of the inset of Fig. 2a starting above the field (*4* gauss) where the slope of the curve in (b) changes.

Fig. 4  The current-voltage characteristics of an array of small junctions that was taken with (a) and without (b) magnetic field shielding. In (b) the earth magnetic field generates an irregular distribution of magnetic flux inside the junctions of the array which ptovides the start-up for the mode observed in the long junctions arrays in zero external magnetic field.

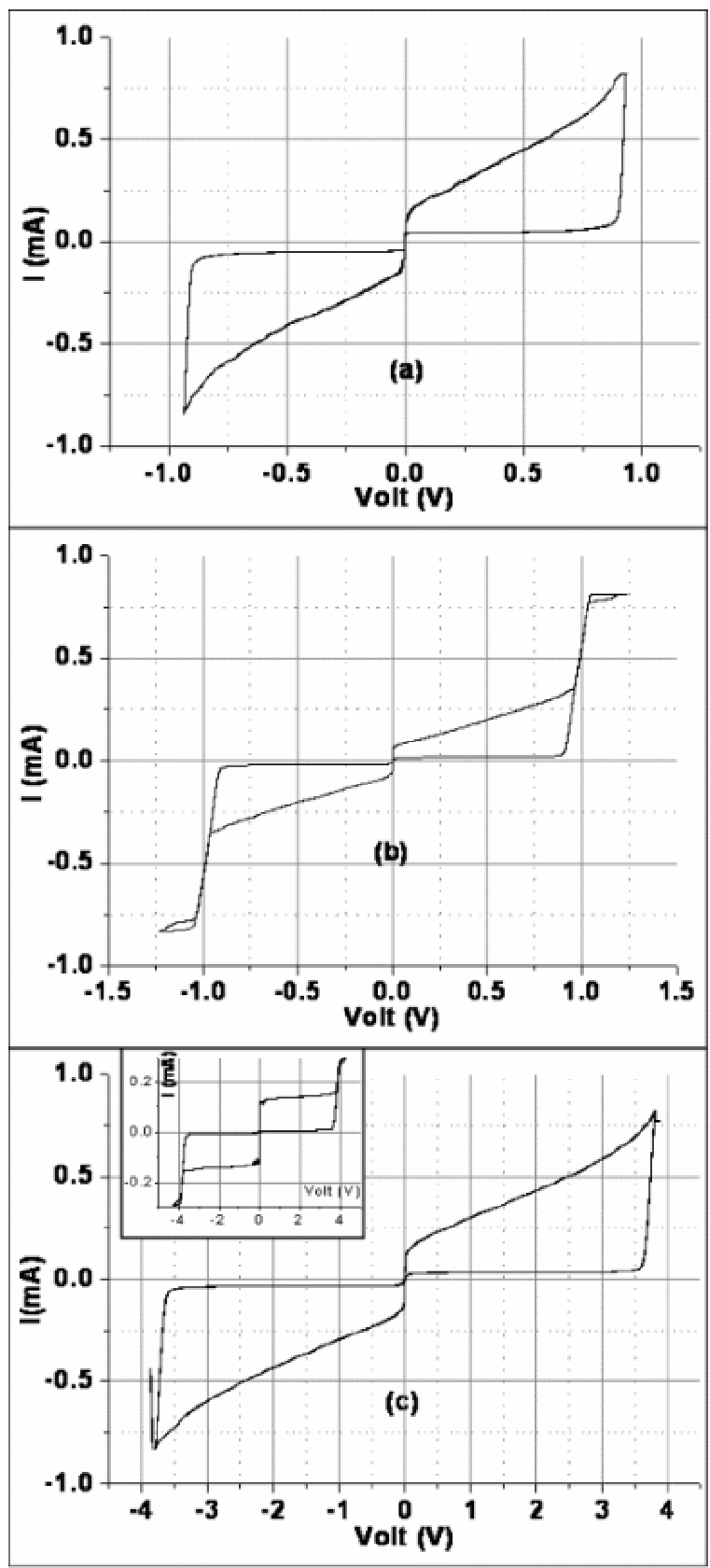

Figure 1, Ottaviani et al.

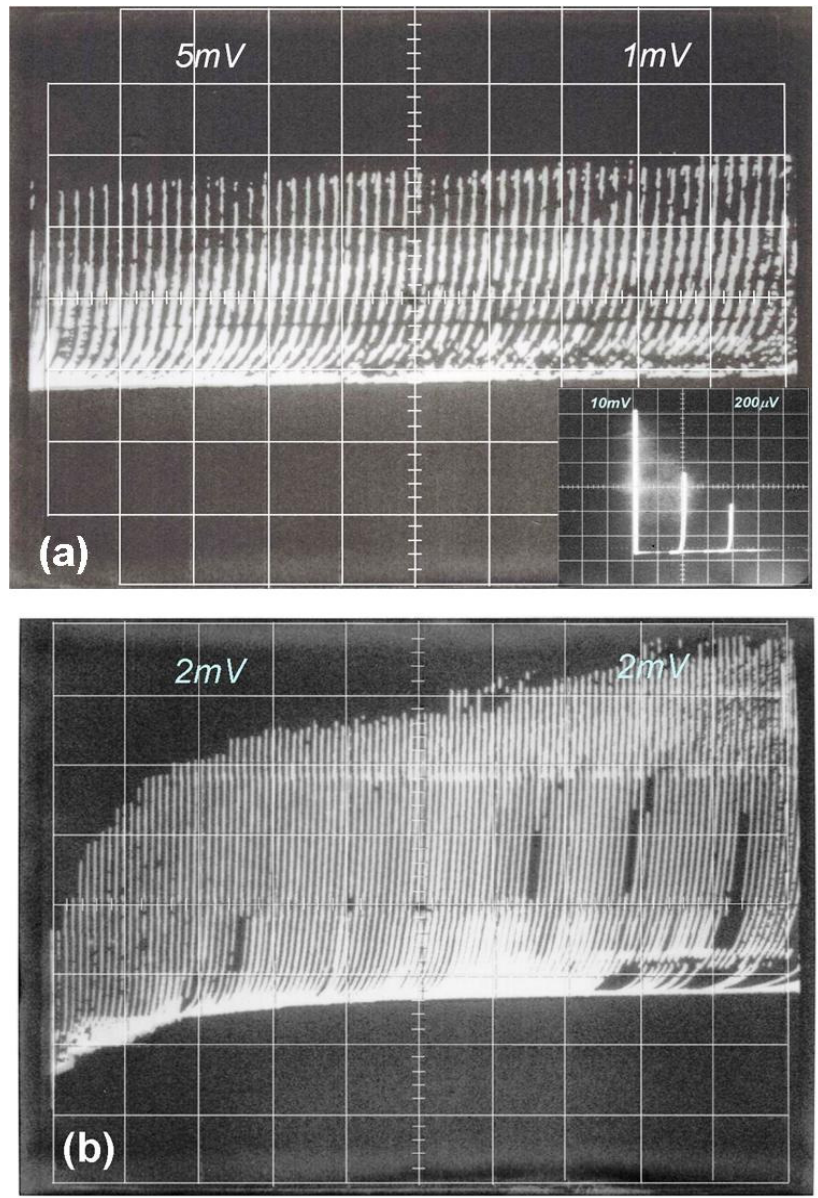

Figure 2, Ottaviani et al.

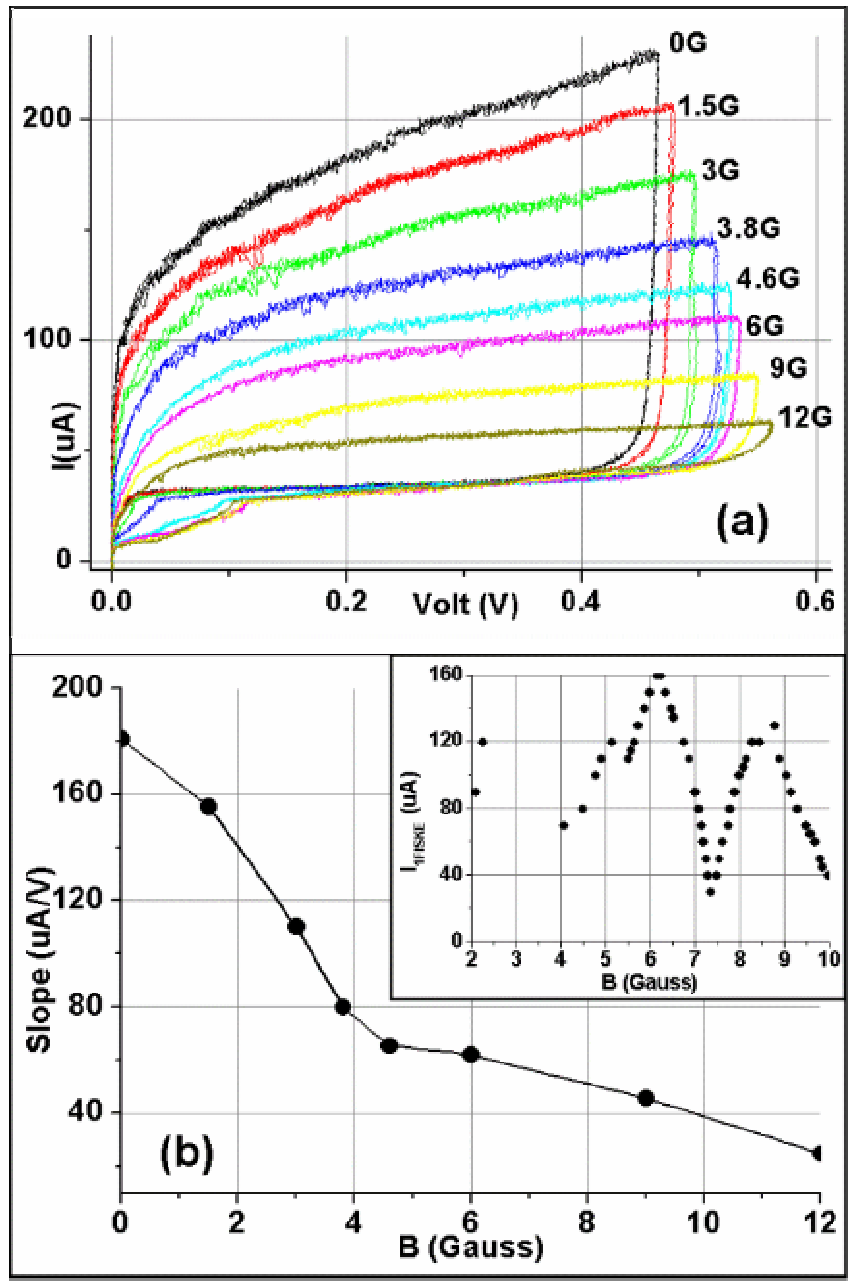

Figure 3, Ottaviani et al.

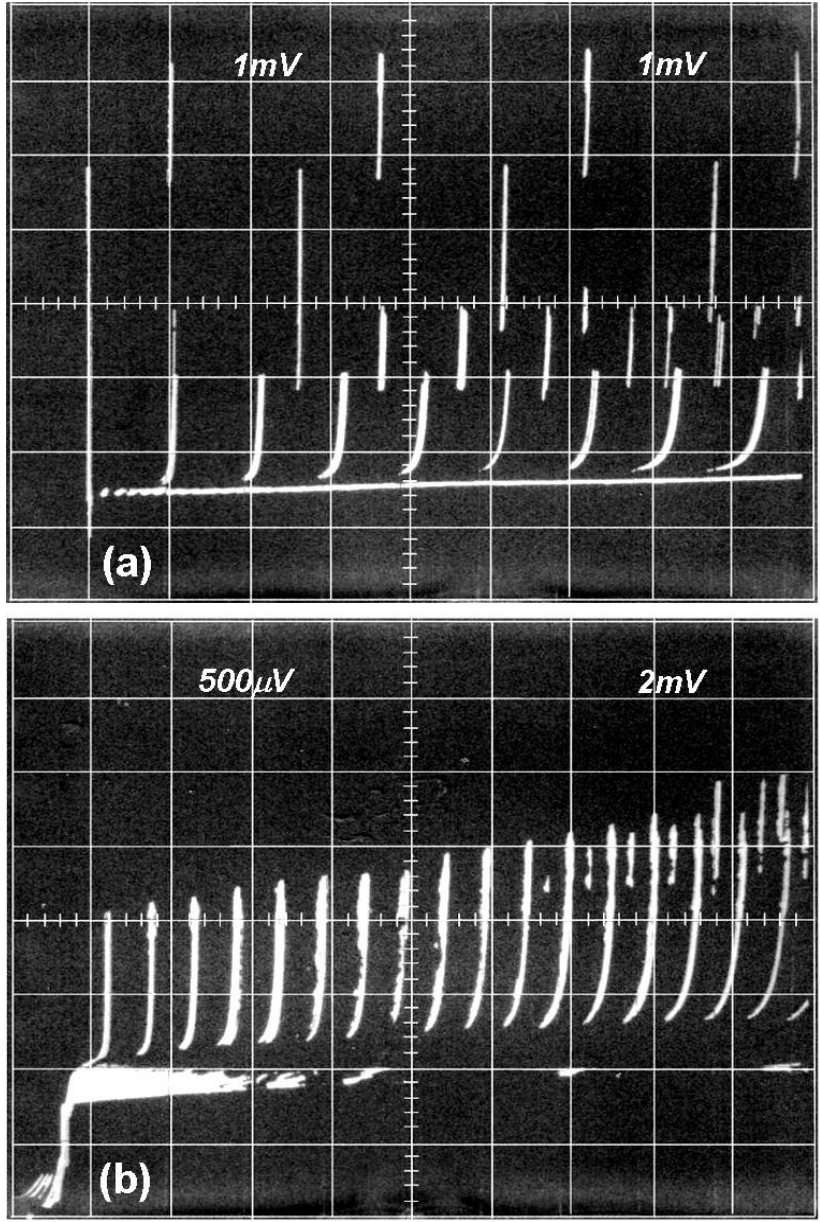

Figure 4, Ottaviani et al.